\documentclass[useAMS,usenatbib]{mn2e}

\def\aap{AA}
\def\aapr{AA Rev}

\def\mnras{MNRAS}
\def\apj{ApJ}
\def\aj{AJ}

\def\pasj{PASJ}
\def\nat{Nat}
\def\arcsinh{{\rm arcsinh}}
\def\threeh{{{3\over 2}}}
\def\half{{{1\over 2}}}
\def\msun{{M_\odot}}
\def\change#1{{#1}}
\def\tb{{\hat b}}
\def\td{{\hat d}}
\def\RDM{{R_{\rm dm}}}
\def\RB{{R_{\rm b}}}
\def\Rthin{{R_{\rm thin}}}
\def\Rthick{{R_{\rm thick}}}
\def\rhoNFW{{\rho_{\rm nfw}}}
\def\argu{{b\sqrt{\chi^2\!+\!q^2\xi^2}}}
\def\ds{\displaystyle}

\usepackage{graphicx}
\usepackage{float}
\usepackage{amssymb}
\usepackage{amsfonts}
\usepackage{amsmath} 
\usepackage{color}
\title[Thick Discs and Haloes: Flattening and Velocity Anisotropy]
      {Virial Sequences for Thick Discs and Haloes: Flattening and
        Global Anisotropy} \author[A. Agnello and
        N. W. Evans]{A. Agnello$^{1}$\thanks{E-mail:
          aagnello@ast.cam.ac.uk,nwe@ast.cam.ac.uk} and
        N. W. Evans$^{1}$\\ $^{1}$Institute of Astronomy, University
        of Cambridge, Madingley Road, Cambridge CB3 0HA, UK}

\begin{document}

\date{Accepted . Received }

\pagerange{\pageref{firstpage}--\pageref{lastpage}} 

\maketitle

\label{firstpage}

\begin{abstract}
  Stellar haloes and thick discs are tracer populations and make only
  a modest contribution to the overall gravity field.  Here, we
  exploit the virial theorem to provide formulae for the ratio of the
  globally-averaged equatorial to vertical velocity dispersion of a
  tracer population in spherical and flattened dark matter haloes.
  This gives {\it virial sequences} of possible physical models in the
  plane of global anisotropy and flattening. The tracer may have any
  density distribution, although there are particularly simple results
  for power-laws and exponentials. We prove the Flattening Theorem:
  for a spheroidally stratified tracer density with axis ratio $q$ in
  equilibrium in a dark density potential with axis ratio $g$, the
  ratio of globally averaged equatorial to vertical velocity
  dispersion depends only on the combination $q/g$. As the stellar
  halo density and velocity dispersion of the Milky Way are accessible
  to observations, this provides a new method for measuring the
  flattening of the dark matter distribution in our Galaxy. If the
  kinematics of the local halo subdwarfs are representative, then the
  Milky Way's dark halo model is oblate with a flattening in the
  potential of $g \approx 0.85$, corresponding to dark matter density
  contours with flattening  $\approx 0.7$.

  The fractional pressure excess for power-law populations is roughly
  proportional to both the ellipticity and the fall-off exponent.
  Given the same pressure excess, if the density profile of one
  stellar population declines more quickly than that of another, then
  it must be rounder.  This implies that the dual halo structure
  claimed by Carollo et al. (2007) for the Galaxy -- a flatter inner
  halo and a rounder outer halo -- is inconsistent with the virial
  theorem.

  For the thick disc, we provide formulae for the virial sequences of
  double-exponential discs in logarithmic and Navarro-Frenk-White
  (NFW) haloes. There are good matches to the observational data on
  the flattening and anisotropy of the thick disc if the thin disc is
  exponential with a short scalelength $\approx 2.6$ kpc and local
  surface density of $56 \pm 6 M_\odot {\rm pc}^{-2}$, together with a
  logarithmic dark halo. Thin discs with long scalelengths $\approx
  3.5$ kpc are disfavoured. Likewise, Navarro-Frenk-White potentials
  do not seem to produce virial sequences matching the thick disc
  kinematics.
\end{abstract}



\section{Introduction}\label{intro}

The steady-state tensor virial theorem (see e.g. Chandrasekhar 1987)
relates the overall geometry of the matter distribution to the global
kinematics.  It is a global result that constrains the components of
the pressure, kinetic energy and potential energy tensors.  The
potential energy tensor $W_{ij}$ depends on the flattening, whilst the
kinetic energy tensor $T_{ij}$ describes the rotational and velocity
anisotropy properties.

Stars and their kinematics provide one of the tools available for
studies of the morphologies and sizes of galaxies and their halos.
So, a classic application of the tensor virial theorem is to the
problem of the shape of elliptical and lenticular galaxies.  Binney
(1978, 2005) gave a pictorial representation of the plane of
ellipticity $\epsilon$ versus ratio of rotational to pressure support
$v/\sigma$ for early-type galaxies. He showed that if elliptical
galaxies have isodensity surfaces that are similar ellipsoids, then
$v/\sigma$ is completely determined by the velocity anisotropy and the
flattening. This diagram is an important tool in understanding whether
the flattening of an elliptical galaxy is due to velocity anisotropy
or rotation.

Although the tensor virial theorem has been widely used in galactic
astronomy, applications have been largely confined to self-gravitating
systems -- such as the stars in bulges or elliptical galaxies ---
which are moving in a potential mainly generated by their own density
field.  However, some populations make a very modest contribution to
the total potential in which they move. An example is the population
of metal-poor halo stars, which are mere tracers of the gravity field
of the overwhelmingly dominant dark halo. Similarly, the stars of the
thick disc move in the combined field of the more massive thin disc
and dark halo. We shall refer to such instances as {\it tracer
  populations}. Here, we provide a systematic investigation of the
tensor virial theorem applied to tracers, including stellar haloes and
thick discs.

Conventional methods of analysis of stellar kinematical data on tracer
populations are usually based upon phase space distribution functions
(e.g, Evans, Hafner \& de Zeeuw 1997, Deason, Evans \& Belokurov
2011a) or upon the Jeans equations (e.g., van der Marel 1991,
Battaglia et al. 2005). The former is most fundamental as it is
equivalent to solving the collisionless Boltzmann equation.  The
method usually depends upon an Ansatz as to the likely phase space
distribution function, with the free parameters fit in a maximum
likelihood sense to the kinematical data. The latter is less
fundamental, as the Jeans equations are moments of the collisionless
Boltzmann equation. The method depends not just on the stellar density
and velocity field, but also explicitly on their gradients, which are
hard to obtain with noisy data and susceptible to deviations caused by
contaminants, interlopers and substructure.

The tensor virial theorem is obtained by integration of moments of the
Jeans equations, and so can be regarded as a globally averaged version
of them. Although there are earlier attempts to use the tensor virial
theorem to study properties of the Galaxy via the kinematics of tracer
stars in the literature (White 1989, Sommer-Larsen \& Christensen
1989), they were held up by a serious drawback. Kinematic quantities,
for example, of halo stars in our Galaxy are only known locally, and
there is a leap of faith required in extrapolating from local to
global quantities.  However, given the many large scale surveys of
Galactic stars underway or planned, it is now worthwhile to examine
the tensor virial theorem anew for its potentiality. Astrometric
satellites such as GAIA (Turon et al. 2005), current spectroscopic
surveys such as SEGUE (Yanny et al. 2009), and forthcoming
initiatives, such as ESO-GAIA and Hermes, offer the prospects of huge
amounts of kinematic information becoming available for many stars
throughout the Galaxy. Understanding these complex and overlapping
datasets, and extracting useful information from the radial velocities
and proper motions, will be a major task over the next few
years. Global averages like the tensor virial theorem may provide a
quick way of extracting ready and accurate information on the
structure of the Galactic potential.

The paper is arranged as follows. In Section 2, we introduce the
tensor virial theorem. Section 3 derives formulae for the ratio of
equatorial to vertical pressure for scale-free tracer populations in
spherical haloes. In Section 4, we develop theorems that extend our
results to the cases of flattened haloes and more general tracer
populations. Section 5 discusses applications to Population II stars
in the stellar halo of the Milky Way. Using observational data on the
pressure ratio, we develop a new method of constraining the flattening
of the dark halo. We also show that virial arguments impose strong
constraints on multiple populations moving in the same potential and
apply our arguments to recent claims of duality in the stellar halo.
Section 6, discusses applications to the thick disc, in particular
focusing on whether the observed kinematics and flattening are
consistent with motion in a thin disc potential with short or long
scalelength. Finally, we revisit the stellar halo in Section 7, asking
whether it is conceivable that the flattening to the potential
provided by the thin disc alone reproduce the observed flattening and
kinematics.

\section{The Tensor Virial Theorem}\label{sect:TVT}

Let $V$ be the volume of our survey. For an all-sky survey like GAIA,
$V$ might be all space. However, this need not always be the case.

Whatever the volume $V$, the kinetic energy tensor $K_{ij}$ is the sum
$2T_{ij} + \Pi_{ij}$, where
\begin{eqnarray}
  T_{ij}& = &\frac{1}{2}\int_V{\rho {\bar v_{i}}
    {\bar v_{j}}} \mathrm{d}^{3}x ,\\
  \Pi_{ij}& = &\int_V{\rho \sigma^{2}_{ij}}\mathrm{d}^{3}x .
\end{eqnarray}
Here, $\rho$ is the stellar density, whilst ${\bar v_i}$ is the
velocity field and $\sigma^{2}_{ij}$ is the velocity dispersion
tensor. Here, an overbar denotes an average over the velocity
distribution so that
\begin{equation}
\sigma^{2}_{ij} = \overline{(v_i - {\bar v_i}) (v_j - {\bar v_j})}.
\end{equation}
We note that $T_{ij}$ is the kinetic energy in ordered, rotational
motion (often negligible for stellar halos but substantial for thick
discs), whilst $\Pi_{ij}$ is the pressure tensor. The potential energy
tensor is
\begin{eqnarray}
  W_{ij} &=& \int_V \rho x_{i}{\partial \Phi \over \partial
    x_{j}} \mathrm{d}^3x\ ,
\end{eqnarray}
where $\Phi$ is the gravitational potential.

To obtain the relationship between these tensors, we start from the
steady-state Jeans equations, which read
\begin{equation} {\partial \over \partial x_k} \rho \overline{v_k v_j}
  = - \rho {\partial \Phi \over \partial x_j}\ .
\end{equation}
Multiplying by $x_i$, integrating over the volume $V$ and making use
of the divergence theorem gives the steady-state, tensor virial
theorem (e.g., Chandrasekhar 1987, Binney 1981)
\begin{equation}
  K_{ij} + W_{ij} = 2T_{ij}+\Pi_{ij}+W_{ij}= S_{ij}\ ,
\end{equation}
where the surface term is
\begin{eqnarray}
  S_{ij} = \oint_{\partial V} x_i \rho \overline{v_j v_k}
  dS_k\ ,
\label{eq:surface}
\end{eqnarray}
and $\partial V$ is the bounding surface.  If the integration volume
$V$ is taken as all space, then the surface term can be discarded
provided that the pressure $\rho \overline{v_jv_k}$ falls off
more quickly than $r^{-3}$.

For an axisymmetric density and potential, we will often find it
convenient to use cylindrical polars ($R, \phi, z$). It is useful to
sum up all the kinetic energy or pressures in the equatorial plane and
refer to
\begin{equation}
K_{RR} = K_{xx}+ K_{yy}, \qquad\qquad \Pi_{RR} = \Pi_{xx}+\Pi_{yy}.
\end{equation}
Analogously, we can also introduce
$$W_{RR} = W_{xx}+W_{yy}\ =\ \int{\rho R \partial_{R}\Phi}\mathrm{d}^{3}x $$
and $S_{RR} = S_{xx}+S_{yy}$.

The great advantage of virial methods is that integrated quantities
(such as total kinetic energy of the stars in the survey volume) can
in general be evaluated robustly even with quite sparse data. A price
to be paid is that the surface terms may not necessarily vanish and
may have to be evaluated from the data.


\section{Tracer Populations in Spherical Dark Haloes}

For the moment, the gravitational potential is assumed to be dominated
by dark matter. The stars comprise a tracer population. This
assumption is fine for stellar haloes which move in the dominating
dark matter potential. For example, the Milky Way Galaxy's stellar
halo has a total mass of $\sim 4 \times 10^8 \msun$ (e.g., Bell et
al. 2008), which is four orders of magnitude smaller than the dark
matter halo (e.g., Watkins et al. 2010).

\subsection{The Pressure Ratio}

Let us take the potential to be spherically symmetric
(i.e. $\Phi(x,y,z)\equiv\Phi(r)$) and the stellar density to be
axisymmetric (i.e., $\rho(x,y,z) \equiv \rho (R,z)$).  Let the
integration volume $V$ be over all space and let the surface terms
vanish, so that the tensor virial theorem can be manipulated to yield:
\begin{equation} {\displaystyle K_{RR} \over \displaystyle K_{zz}} 
  =\ {\displaystyle \int { \rho R^2 \Phi^\prime \over \displaystyle r}
     \mathrm{d}^3x \over 
    {\displaystyle \int {\rho z^2 \Phi^\prime \over \displaystyle r}\mathrm{d}^3x}}\ , 
\label{eq:TVTratio}
\end{equation}
where $\Phi^\prime = \mathrm{d}\Phi /\mathrm{d}r$.  The quantity
${K_{RR} / K_{zz}}$ is the ratio of the equatorial or in-plane kinetic
energy to the vertical kinetic energy.

Stellar haloes are only weakly rotating (e.g., Deason et al. 2011a),
so it is to fair to assume that there is no mean streaming motion
($T_{ij}=0$). Then, ${K_{RR} / K_{zz}}$ is the same as ${\Pi_{RR} /
  \Pi_{zz}}$, the ratio of pressure in the equatorial to vertical
directions. For an axisymmetric figure, it is natural to expect this to
be intimately related to the flattening.

\subsection{Scale-Free Spheroidal Densities}

To begin with, let us assume that the tracer density is scale-free and
has the form
\begin{equation}
\rho = {f(\theta)\over r^\gamma}
\end{equation}
where ($r,\theta$) are spherical polars and $\gamma$ is the power-law
fall-off. Formally, the scale-free density is non-integrable either at
the origin or at infinity and so the virial quantities do not
exist. Nonetheless, their ratio is well-defined and given by
\begin{equation}
  {\ds W_{RR} \over \ds W_{zz}} = {\ds \Pi_{RR} \over \ds \Pi_{zz}} =
  { \ds \int_0^\pi
    f(\theta)\sin^3 \theta \mathrm{d} \theta \over \ds \int_0^\pi
    f(\theta) \sin \theta \cos^2 \theta \mathrm{d}\theta}\ .
\label{eq:scalefreea}
\end{equation}
As an illustration, we specialise to the case in which the tracer
density is stratified on similar concentric spheroids as
\begin{equation}  
\rho=\rho_{0}\left(R^{2}+z^{2}q^{-2} \right)^{-\gamma/2},
\end{equation}
with ($R,\phi,z$) familiar cylindrical polar coordinates.  Here, $q$
is the axis ratio of the equidensity contours, which is simply related
to the ellipticity $\epsilon = 1-q$. The virial ratio
eq~(\ref{eq:scalefreea}) is
\begin{equation}
  {\ds W_{RR} \over \ds W_{zz}} = {\ds \Pi_{RR} \over \ds \Pi_{zz}} =
  {\ds \int_0^\pi \sin^3 \theta (\sin^2 \theta + q^{-2} \cos^2
    \theta)^{-\gamma/2} \mathrm{d} \theta \over \ds \int_0^\pi \sin \theta
    \cos^2 \theta (\sin^2\theta + q^{-2} \cos^2
    \theta)^{-\gamma/2}\mathrm{d} \theta}\ ,
\label{eq:crux}
\end{equation}
which is independent of $\Phi$

Some simple and useful results can be obtained in the case of modest
flattening, $q\rightarrow1$. Using the shorthand $Q = q^{-2}$, then
Taylor expansion yields
\begin{equation}
  {\ds \Pi_{RR} \over \ds \Pi_{zz}} = {\ds \int_0^\pi \sin^3 \theta (1
    - \half \gamma [Q-1] \cos^2 \theta ) \mathrm{d} \theta \over \ds 
    \int_0^\pi \sin \theta \cos^2 \theta ( 1- \half \gamma [Q-1]
    \cos^2\theta) \mathrm{d}\theta}.
  \label{eq:cruxsimple}
\end{equation}
We can get an expression for the vertical and horizontal pressure
difference by making the definitions
$\langle\sigma^{2}_{zz}\rangle=\Pi_{zz}/M$ and
$\langle\sigma^{2}_{xx}\rangle= \frac{1}{2}\Pi_{RR}/M$, where $M$ is
the total mass of the tracer population.  The angled brackets
therefore denote mass-weighted averages of the velocity
dispersions. We can now recast eq~(\ref{eq:cruxsimple}) as
\begin{equation} 
  \frac{\langle\sigma^{2}_{xx}\rangle}{\langle\sigma^{2}_{zz}\rangle}\ \approx
1 + {\gamma \over 5} (Q-1)  + {3\gamma (\gamma-5) \over 175} (Q-1)^2.
\end{equation}
Bearing in mind that $Q = 1/q^2 = 1/(1-\epsilon)^2$, where $\epsilon$
is the ellipticity, we see that the fractional pressure excess is
\begin{equation}
  \frac{\langle\sigma^{2}_{xx}\rangle - \langle\sigma^{2}_{zz}\rangle}
{\langle\sigma^{2}_{zz}\rangle}\ \approx {1\over 5} \gamma (Q-1)
\approx {2\over 5} \gamma \epsilon \ .
\label{eq:figures}
\end{equation}
In other words, for a mildly flattened stellar halo with a given
ellipticity and power-law density fall-off residing in a spherical
dark halo, then
\begin{equation} {\rm Fractional\ pressure\ excess} \approx {2\over 5}
  \times {\rm Falloff} \times {\rm Ellipticity}\ .
\label{eq:words}
\end{equation}
We note that, when
$\langle\sigma^{2}_{xx}\rangle\rightarrow\langle\sigma^{2}_{zz}\rangle,$
$Q$ and hence $q$ tend to $1$ (i.e. the stellar halo is spherically
stratified), and vice-versa. This behaviour in the isotropic pressure
limit is general for spherical dark matter haloes.

In the limit of extreme flattening, an asymptotic expansion of
eq~(\ref{eq:crux}) yields
\begin{equation}
\frac{\Pi_{RR}}{\Pi_{zz}}\sim
\begin{cases}
(\gamma-3)q^{-2} & \text{if $\gamma>3$}\\
q^{-2}/\log(2q^{-2})& \text{if $\gamma=3$}\\
\frac{1}{2}(3\!-\!\gamma)q^{1-\gamma}B(\frac{\gamma\!-\!1}{2},\frac{\gamma}{2}) & \text{if
  $1<\gamma<3$}
\end{cases}
\label{eq:highlyflat}
\end{equation}
with $B(a;b)=\Gamma(a)\Gamma(b)/\Gamma(a+b)$ being the Beta
function. Stellar halo populations typically have $\gamma \gtrsim 3$
and so the pressure ratio $\Pi_{RR}/\Pi_{zz} \sim q^{-2}$ for high
flattening. We will see later in Section~\ref{sec:TD} how the presence
of a thin disc alters this behaviour.

\begin{figure}
        \centering
\includegraphics[width=0.45\textwidth]{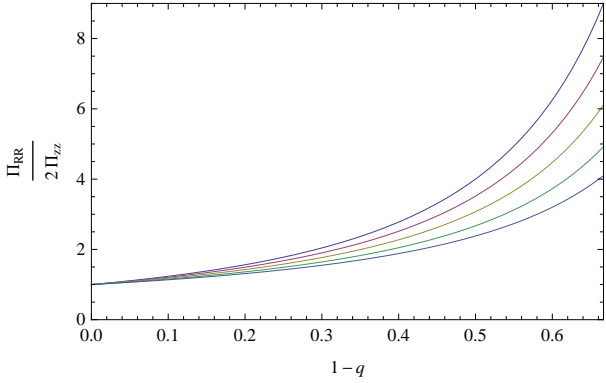}
\caption{Virial sequences for spheroidal stellar halos. The pressure
  ratio $\half \Pi_{RR}/ \Pi_{zz}$ is shown as a function of the
  flattening $q$ for stellar densities with fall-off $\gamma =3$
  (blue), 3.5 (green), 4 (yellow), 4.5 (red) and 5 (purple). The
  pre-factor of $\half$ means that the vertical axis is $\Pi_{xx}/
  \Pi_{zz}$ and ensures that the ratio tends to unity as $q
  \rightarrow 1$. The analytic form of these curves is given in
  eqs~(\ref{eq:casem3}), (\ref{eq:casem4}) and (\ref{eq:casem5}) for
  $\gamma =3$, 4 and 5.}
\label{fig:power}
\end{figure}

\subsection{Exact Results}

In fact, the pressure ratio in eq~(\ref{eq:crux}) is analytic in the
case that power-law fall-off $\gamma$ is an integer. It is helpful to
have the exact expressions.  For $\gamma =2$, we find:
\begin{equation} {\Pi_{RR} \over \Pi_{zz}}= \ \frac{Q \arctan
    \sqrt{Q-1} - (Q-1)^\half}{(Q-1)^\half - \arctan \sqrt{Q-1}}.
\end{equation}
For the case $\gamma =3$, we have:
\begin{equation}
  \frac{\Pi_{RR}}{\Pi_{zz}}\ =
  \ \frac{(Q-1)^{\threeh} - \sqrt{Q} \arcsinh\sqrt{Q-1} + (Q-1)^\half}
  {\sqrt{Q} \arcsinh \sqrt{Q-1} - (Q-1)^\half}\ .
\label{eq:casem3}
\end{equation}
This formula is equivalent to one derived by White (1989) for
populations orbiting in an isothermal sphere.

The $\gamma =4$ case has already been discussed by Helmi (2008). We
find:
\begin{equation}
  \frac{\Pi_{RR}}{\Pi_{zz}}\
  =\ \frac{ Q(Q-2)\arctan \sqrt{Q-1} + (Q-1)^\half}{Q\arctan
    \sqrt{Q-1} - (Q-1)^\half}\ .
\label{eq:casem4}
\end{equation}
Like us, Helmi assumes that the stellar halo is flattened and follows
a scale-free power-law density, whilst the dark halo is spherical. In
fact, Helmi specialises to the particular case of a logarithmic
potential generating an asymptotically flat rotation curve.  To make
progress, Helmi makes an additional assumption, namely that the ratio
of horizontal to vertical velocity dispersions
$\sigma_{x}^{2}/\sigma_{z}^{2}$ does not depend on location.
Unfortunately, the final formula -- equation (8) of Helmi (2008) --
contains two mistakes. The first is computational, namely the
integrals are incorrect. This can be seen by noting that, in the limit
of a spherical stellar halo ($Q \rightarrow 1$), the velocity
dispersion tensor must become isotropic within Helmi's set of
assumptions, a fact that is contradicted by her formula. The second is
more subtle and conceptual. The ratio $\sigma_{x}^{2}/\sigma_{z}^{2}$
between velocity dispersions in the $x$ and $z$ coordinates cannot be
considered uniform, once the density and potential profiles are
already assigned: this additional assumption over-determines the
problem.

For the $\gamma =5$ case, the result is exceptionally simple, namely
\begin{equation}
  \frac{\Pi_{RR}}{\Pi_{zz}}\
  =2Q = {2\over q^2}.
\label{eq:casem5}
\end{equation}
The result is exact for arbitrary flattening.

Figure~\ref{fig:power} shows the behaviour of $\Pi_{xx}/\Pi_{zz}$
valid for any spherical potential.  It assumes a power-law profile for
the stellar density stratified on similar concentric spheroids with
axis ratio $q$ and with density fall-off $m$ varying between 3 (lower
curve) and 5 (upper curve). Populations in the stellar halo typically
fall off like $r^{-3}$ or $r^{-4}$, though some variable star
populations such as RR Lyrae (e.g., Watkins et al. 2009) and Miras
(e.g., An et al. 2004) can fall off faster.  As a general rule, the
faster the density fall-off, and the greater the flattening, then the
larger the pressure excess needed to support the stellar distribution.

\subsection{Alignment of the Velocity Ellipsoid}

To make practical use of the virial formulae, we must find the
connection between the pressure ratio and the velocity
dispersions. This is straightforward if the velocity dispersion tensor
is diagonal in cylindrical polar coordinates, namely
\begin{equation}
{\Pi_{RR} \over \Pi_{zz} }
=  {\ds\int\rho(\sigma^{2}_{RR} + \sigma^2_{\phi\phi})\mathrm{d}^3x
  \over \ds \int{\rho
    \sigma^{2}_{zz}\mathrm{d}^3x}}\ .
\end{equation}

If the velocity dispersion tensor is aligned in spherical polar
coordinates -- as is the case for halo stars (e.g., Smith et
al. 2009b) -- then matters are more complicated. The diagonal
components of the velocity dispersion tensor in cylindrical polar
coordinates are now given by
\begin{subequations}\label{eq:sigt}\begin{align}
\sigma_{RR}^2&=\sigma_{rr}^2\sin^2\!\theta+\sigma_{\theta\theta}^2\cos^2\!\theta
=(1-\beta\cos^2\!\theta)\,\sigma_{rr}^2
\\\sigma_{\phi\phi}^2&=\sigma_{\theta\theta}^2=(1-\beta)\,\sigma_{rr}^2
\\\sigma_{zz}^2&=\sigma_{rr}^2\cos^2\!\theta+\sigma_{\theta\theta}^2\sin^2\!\theta
=(1-\beta\sin^2\!\theta)\,\sigma_{rr}^2
\end{align}\end{subequations}
where $\beta$ is the anisotropy parameter,
\begin{equation}
  \beta=1
  -{\sigma_{\theta\theta}^2 \over \sigma_{rr}^2}\ .
\label{eq:betadef}
\end{equation}
This means that
\begin{equation}
{\Pi_{RR} \over \Pi_{zz} }
=  {\ds \int\rho\sigma^{2}_{rr}( 2- {\beta} + {\beta}
  \cos^2 \theta) \mathrm{d}^3x \over \ds \int{\rho
    \sigma^{2}_{rr}( 1 - \beta \sin^2 \theta) \mathrm{d}^3x}}\ .
\end{equation}
If $\beta$ is constant, this simplifies to
\begin{equation}
{\Pi_{RR} \over \Pi_{zz} } = { \ds (2 - \beta) \langle \sigma^2_{rr} \rangle
- \beta \langle \sigma^2_{rr} \cos^2 \theta \rangle \over \ds  \langle
\sigma^2_{rr} \rangle - \beta \langle \sigma^2_{rr} \sin^2\theta \rangle}
\label{eq:anisotropy}
\end{equation}
\change{where angled brackets denote mass-weighted averages. If the
  tracer flattening is known and the variation of the velocity
  dispersions is estimated over the survey volume,
  eq~(\ref{eq:anisotropy}) yields virial sequences that do
  \textsl{not} involve the potential $\Phi$. Thus, we can constrain
  the potential by intersecting the $W_{RR}/W_{zz}$ sequences with the
  pressure ratios.  Nonetheless, practical application of
  eq~(\ref{eq:anisotropy}) is limited by the fact that the global
  variation of the dispersions with position is not well understood,
  though this will be remedied after the next generation of
  large-scale surveys of the Milky Way (such as GAIA).}

\begin{figure*}
        \centering
\includegraphics[width=0.45\textwidth]{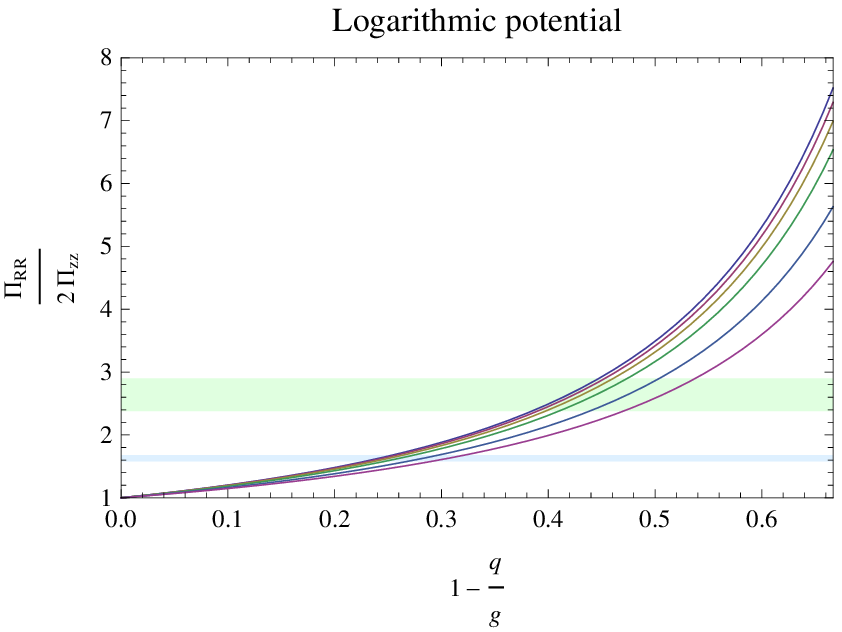}
\includegraphics[width=0.45\textwidth]{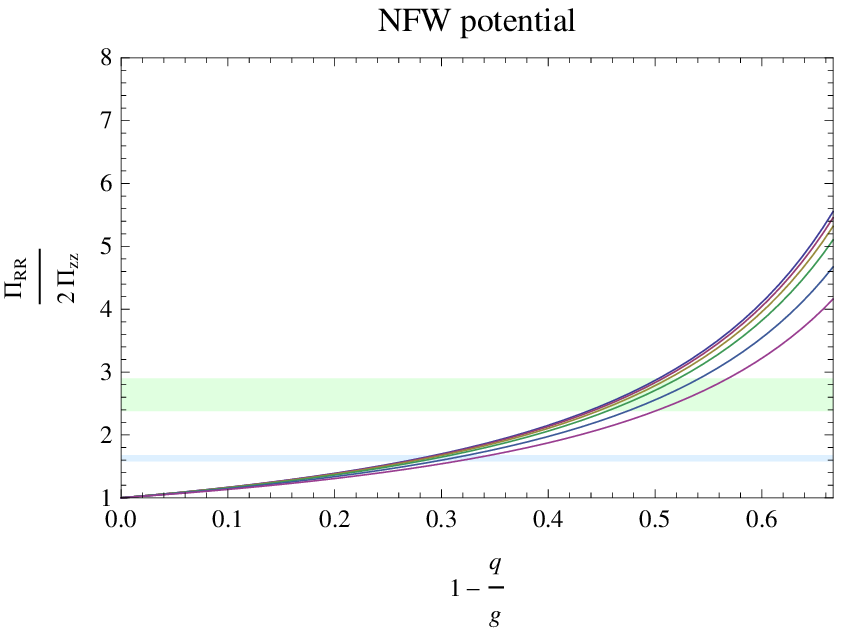}
\caption{Virial sequences for the stellar halo density law found by
  Deason et al. (2011b) and given in eq.~(\ref{eq:densAlis}). The
  pressure ratio $\half \Pi_{RR}/ \Pi_{zz}$ is shown as a function of
  flattening ratio $q/g$, where $q$ is the axis ratio of the stellar
  density contours and $g$ is the axis ratio of the dark matter
  equipotentials. The dark matter potential has a logarithmic form
  generating a flat rotation curve (left) and a NFW dark matter halo
  (right). The coloured lines refer to different values of the dark
  matter scalelength from $5$ kpc (lower red curve) through 15 kpc, 35
  kpc, 55 kpc, 75 kpc to $95$ kpc (upper blue curve) and taking the
  break radius fixed at $\RB = 27$ kpc (Deason et al. 2011b). The green
  and blue bands give the range in pressure ratios found by global
  extrapolation of the local results of Kepley et al. (2007) and by
  Smith et al. (2009a).}
\label{fig:AlisLaw}
\end{figure*}

\section{Two Theorems}

\subsection{The Flattening Theorem}

If our formulae were restricted to spherical dark haloes, they would
have limited applicability. Here, we derive a remarkable theorem that
extends our results to the case of flattened tracer populations in
flattened dark haloes.

Let the tracer density and total potential satisfy the following
requirements
\begin{equation}
\rho(R,qz)=\rho(qz,R),\qquad\qquad \Phi(R,gz)=\Phi(gz,R),
\end{equation}
for some $q$ and $g$. We say in this case that $\rho$ and $\Phi$ are
stratified on constant flattening surfaces. The above definition is
equivalent to requiring that $\rho$ (or $\Phi$) be a symmetric
function of $R$ and $z/q$ (or $z/g$). The most commonly used examples
are spheroidal distributions, such as
\begin{equation}
\Phi=\Phi(m), \qquad\qquad m^2 = R^{2}+ z^{2} g^{-2}.
\end{equation}
There are a number of popular halo models of this form, including
Binney's (1978) logarithmic potential and the power-law models (Evans
1993, 1994). The dark matter density may be cored or cusped. The
flattening, or the minor to major axis ratio, of the dark matter
equipotentials $g$ is rounder than the dark matter isodensity
contours. Similarly, the spheroidal power-law densities with fall-off
$\gamma$ and axis ratio $q$ studied in the previous section belong to
this class, though there are many other possibilities for the tracer
density as well.

In what follows, it is convenient to specialize to spheroidally
stratified potentials with flattening $g$. The tensor virial theorem
becomes:
\begin{equation} {\ds \Pi_{RR} \over \ds \Pi_{zz}} = 
  \ {\displaystyle \int { \rho R\partial_{R} \Phi}
    \mathrm{d}^3x \over 
    {\displaystyle \int {\rho z\partial_{z} \Phi}\mathrm{d}^3x}}\ =
  \ {\displaystyle \int {\rho R^{2} \Phi^{\prime} \over m}
    \mathrm{d}^3x \over 
    {\displaystyle \int {\rho g^{-2}z^{2}\Phi^{\prime} \over m}\mathrm{d}^3x}}\ ,
\label{eq:TVTrationew}
\end{equation}
where $\Phi^\prime = \mathrm{d}\Phi /\mathrm{d}m$. 
 
Making the substitution $z = z^{\prime}g,$ writing $r^{\prime
  2}=x^{2}+y^{2}+z^{\prime 2}$ and then dropping the primes, the
integrals on the right-hand side of eq~(\ref{eq:TVTrationew}) are
converted into the exact form as those on the right-hand side of
eq~(\ref{eq:TVTratio}), with the only change
\begin{equation} 
\rho(R, z/q) \rightarrow 
\rho(R, g z/q ).
\end{equation}
This means that all our preceding results for spherical haloes hold
good in the flattened case, but with the replacement
\begin{equation}
q \rightarrow q/g.
\end{equation}
This considerably extends the scope of our earlier results.

More formally, our theorem can be stated as follows. {\it If the dark
  matter potential $\Phi$ is homoeoidally stratified with flattening
  $g$ and the tracer density $\rho$ has a constant flattening $q$,
  then the virial ratio $W_{RR}/W_{zz}$ depends on the two flattenings
  just through their ratio $q/g$.}

As a corollary, for scale-free densities with constant flattening the
virial ratios depend on $q/g$ and the power-law exponent, but
\textsl{not} on the radial form of the potential, as already noticed
for spherical potentials.  This can be seen by using spherical polar
coordinates and noting that the radial integrals (containing $\Phi$)
factorize and cancel top and bottom. The angular integrals do depend
on the density stratification and the behaviour of $W_{RR}/W_{zz}$,
and thence $\Pi_{RR}/\Pi_{zz}$, is controlled by those:
\begin{equation} { \Pi_{RR} \over \Pi_{zz}} =\
  {\ds \int{\rho(\sin\theta,(g/q)\cos\theta)
      \sin^{3}\theta}\mathrm{d}\theta \over \ds
      \int{\rho(\sin\theta,(g/q)\cos\theta)
      \cos^{2}\theta\sin\theta}\mathrm{d}\theta}\ .
\end{equation}
As a consequence, once a virial ratio has been computed for a
spherical dark matter potential, the flattened analogue follows in a
straightforward manner.  Providing the potential remains spheroidally
stratified, any halo model such as isothermal or Navarro-Frenk-White
(1998) or Sersic (1963) is fine. This is an extension of a famous and
elegant result from potential theory due to Roberts (1962), which
states that ratios such as $W_{RR}/W_{zz}$ depends only on the axial
ratios for homoeoidal distributions, and not on the radial profiles.

The ratio $W_{RR}/W_{zz}$ does depend on the form of $\Phi$, if more
complex profiles are prescribed, for example, accounting for the
contribution of the stellar disc, or the differential flattening of
the dark matter halo and the self-gravity of the stellar halo. 
However, although analytic progress is difficult for complex profiles,
the relevant integrals are easily evaluated numerically.

\subsection{The Conjugateness Theorem}

As shown above, there is a class of test densities (like scale-free
ones) yielding virial ratios that are independent of the radial
behaviour of $\Phi$. Similarly, we can ask whether there are
potentials giving virial ratios that are independent of the radial
stratification of $\rho$. If a density and a potential belong to these
classes and give the same virial ratios, we refer to them as
\textit{conjugate}. We shall demonstrate that scale-free potentials
are conjugate to scale-free tracer densities.

Consider a scale-free power-law potential describing the dark matter
\begin{equation}
  \Phi(r) = \begin{cases} {\displaystyle v_0^2 \over \displaystyle
      \alpha r^\alpha} & \text{if $\alpha \neq 0$}, \\ \null & \null
    \\ v_0^2 \log r & \text{if $\alpha =0$}
\end{cases}
\end{equation}
where $\alpha$ is a constant. Now, the circular velocity in the
equatorial plane varies like $r^{-\alpha/2}$. So models with $\alpha
<0$ have falling rotation curves, models with $\alpha >0$ have rising
rotation curves. When $\alpha =0$, the model is the familiar
isothermal sphere with a perfectly flat rotation curve $v_0$.

Now suppose that the tracer density has the form
\begin{equation}
\qquad \rho = \rho \left ( {r \over h(\theta)} \right).
\end{equation}
Note that this density profile is not necessarily scale-free, but
could be of exponential or Gaussian form. Then, the ratio of
equatorial to vertical pressure is
\begin{equation}
  {\Pi_{RR} \over \Pi_{zz}} = {\ds \int_0^\pi [h(\theta)]^{3 + \alpha} \sin^3 \theta
    \mathrm{d} \theta \over \ds \int_0^\pi  [h(\theta)]^{3+
      \alpha}\sin \theta \cos^2 \theta \mathrm{d} \theta}\ .
\label{eq:scalefreeb}
\end{equation}
But, this is the same virial ratio as would be given by
\begin{equation}
\Phi = \Phi(r), \qquad \rho = {h(\theta) \over r^\gamma}
\end{equation}
with $\gamma = \alpha +3$. The two combinations are conjugate.

The conjugateness theorem then reads: {\it A scale-free tracer density
  with fall-off $\alpha+3$ in an arbitrary spherical potential is
  conjugate to a tracer density in a scale-free potential with
  fall-off $\alpha$.}  To give a practical application of this result,
we might want to know the pressure ratio of a tracer population with
\begin{equation}
\rho = \rho_0 \exp \left[ -\left( R^2 + z^2 q^{-2}
  \right)^{1/2}/\Rthick \right]
\end{equation}
in a galaxy with a flat rotation curve ($\alpha =0$). The density law
might describe a thick disc with horizontal scalelength $\Rthick$ and
vertical scaleheight $q\Rthick$.  By the conjugateness theorem, this
is just the same as the pressure ratio for a spheroidal population
with $\gamma = 3$, which is given in eq~({\ref{eq:casem3}).

We finally remark that, although the potential and the density may be
chosen as scale-free, they cannot both be so chosen as the virial
ratios are then undefined. White's (1989) results for spheroidal
tracer populations orbiting in the singular isothermal sphere hold
good providing the core radius of the tracers $R_{\rm c}$ does not
vanish. Sommer-Larsen \& Christensen (1989) give formulae for the
virial ratios of scale-free populations in scale-free potentials, in
which the core radius of the dark matter $\RDM$ is also zero.
However, the virial ratios depend on the manner in which the limit
$R_{\rm c} \rightarrow 0$ and $\RDM \rightarrow 0$ is approached, and
so their results must be treated with caution.

More precisely, given a power-law density and potential with
core-radii $R_{\rm c}$ and $\RDM$ respectively, the virial ratios can
be easily shown to depend just on the ratio $R_{\rm c}/\RDM$. The
limits $R_{c} \rightarrow 0$ and $\RDM\rightarrow0$ can be approached
by keeping the ratio fixed. Since different $R_{\rm c}/\RDM$ give
different virial ratios, the final limit depends on the path by which
the limit is approached, and so is not defined.

\begin{table}
\begin{center}
\renewcommand{\tabcolsep}{0.2cm}
\renewcommand{\arraystretch}{0.5}
\begin{tabular}{| l l  l  l  l |}
    \hline 
Dataset & Dark Halo & $\Pi_{RR}/\Pi_{zz}$ &  $q/g$ & $g$\\
& Model & & &     \\
    \hline
    \\
K07 & NFW    & $5.28 \pm 0.52$ & $0.52 \pm 0.06$ & $1.2 \pm 0.14$ \\
    \\
K07 & Logarithmic    & $5.28 \pm 0.52$ & $0.55 \pm 0.06$ & $1.1 \pm 0.12 $\\
    \\ \hline
    \\
S09 & NFW    & $3.26 \pm 0.10$ & $0.70 \pm 0.03$ & $0.85 \pm 0.04$  \\
    \\ 
S09 & Logarithmic  & $3.26 \pm 0.10$ & $0.73 \pm 0.03$ & $0.82 \pm 0.04$   \\
    \\ 
    \hline
  \end{tabular}
  \caption{Results for the ratio of stellar halo to dark halo
    flattenings $q/g$ for different assumed halo models using the
    Kepley et al. (2007) and the Smith et al. (2009a) datasets. The
    value of the dark halo axis ratio $g$ is given assuming that the
    stellar halo flattening is $q=0.6$ as found by Deason et
    al. (2011b).}
\label{tab:halores}
\end{center}
\end{table}

\section{Stellar Haloes} 

\label{sec:apps}

\subsection{The Flattening of the Milky Way Dark Halo}

The flattening in the stellar density $q$ and the pressure ratio
$\Pi_{RR} /\Pi_{zz}$ are in principle accessible directly from
starcounts and kinematical measurements of halo stars. The flattening
in the dark halo density $g$ is not. The Flattening Theorem therefore
provides a completely new method to study the dark halo flattening.

Deason et al. (2011b) traced the stellar halo of the Milky Way using
blue stragglers and blue horizontal branch stars as tracers. They found
that the halo has a broken density profile given by
\begin{equation}
\rho(R,z)=\left\{
\begin{array}{l}
\rho_{0}\left(\frac{\sqrt{R^{2}+z^{2}/q^{2}}}{\RB}\right)^{-2.35}\ ,\
\sqrt{R^{2}+z^{2}/q^{2}} \leq \RB,\\
\rho_{0}\left(\frac{\sqrt{R^{2}+z^{2}/q^{2}}}{\RB}\right)^{-4.6}\ ,\
\sqrt{R^{2}+z^{2}/q^{2}} \geq \RB.
\end{array}
\right.
\label{eq:densAlis}
\end{equation}
This law has an explicit break radius $\RB$, at which the
power-law profile changes from a shallow inner slope of 2.35 to a
steeper outer one of 4.6. Deason et al. (2011b) presented evidence
that $\RB = 27$ kpc and that $q = 0.6$.

Fig.~\ref{fig:AlisLaw} shows curves of pressure ratio for the stellar
density law~(\ref{eq:densAlis}) as a function of $q/g$, where $q$ is
the axis ratio of the stellar density contours, whilst $g$ is the axis
ratio of the dark matter equipotentials.  Now, the density law is no
longer a homogeneous function of coordinates, and so our results do
depend on the radial profile of the gravitational potential. The left
panel gives results for a logarithmic potential
\begin{equation}
\Phi = {v_0^2\over 2} \log (\RDM^2 + m^2), \qquad\qquad m^2= R^2 + z^2g^{-2}.
\label{eq:binneysmodel}
\end{equation}
The right panel shows the outcome for the same stellar density, but
now in a spheroidally stratified Navarro-Frenk-White potential
\begin{equation}
  \Phi = -{4\pi \rhoNFW \RDM^3\over m } \log \left( 1+ {m \over \RDM} \right),
  \label{eq:nfwmodel}
\end{equation}
which has an everywhere positive density providing $g> 1/\sqrt{2}$.
The different coloured lines show the effect of varying $\RDM$ between
$5$ kpc (lower curve) and $95$ kpc (upper curve), using a fixed break
radius $\RB = 27$ kpc as advocated by Deason et al. (2011b).

The stellar kinematics of halo stars in the local neighbourhood have
been measured many times. Specifically, we assume that the locally
calculated ratio of stellar velocity dispersions is a good
approximation to the density-weighted globally averaged ratio. Even
though the velocity dispersions of course change with position, the
local ratio can still be an excellent approximation to the global one
provided the anisotropy does not vary dramatically with
position. \change{Notice that we are not explicitly assuming either
spherical or cylindrical alignment (which in any case coincide
locally), although we are arguably making a stronger assumption in
taking local ratios as representative of global ones. Of course, this
limitation to our calculation will vanish once surveys such as GAIA
have been completed and provide a global picture of how the velocity
dispersion varies for components in the Galaxy.}

Kepley et al. (2007) assembled a sample of 221 halo stars within 2.5
kpc of the Sun with distances, radial velocities and proper
motions. They found $(\sigma_{RR}, \sigma_{\phi\phi}, \sigma_{zz}) =
(175 \pm 8, 110 \pm 6, 84 \pm 4)$ kms$^{-1}$ around means of $(\bar
v_R, \bar v_\phi, \bar v_z ) = ( -4 \pm 11, -23 \pm 8, -1 \pm 6$)
kms$^{-1}$. This gives $\Pi_{RR} / \Pi_{zz} = 5.28 \pm 0.52$, which is
represented as the green horizontal band in Fig.~\ref{fig:AlisLaw}.
As the Kepley et al. data are drawn from a variety of disparate
sources, the errors are appreciable. Smith et al. (2009a) used a sample
of $\approx 1700$ halo subdwarfs extracted from the light-motion curve
catalogue of Bramich et al. (2008) of Sloan Digital Sky Survey (SDSS)
Stripe 82.  The halo subdwarfs were extracted using a reduced proper
motion diagram.  This is a larger and more homogeneous sample of halo
stars centered on $R= 8.70$ kpc and $z = -2.41$ kpc, and so below the
Galactic plane. Smith et al. found $(\sigma_{RR}, \sigma_{\phi\phi},
\sigma_{zz}) = (138 \pm 2, 82 \pm 1, 89 \pm 1)$ kms$^{-1}$ around
means of $(\bar v_R, \bar v_\phi, \bar v_z ) = ( 9 \pm 3, 2 \pm 2, -1
\pm 2$) kms$^{-1}$. This gives $\Pi_{RR}/ \Pi_{zz} = 3.26 \pm 0.10$,
which is shown as a light blue horizontal band on
Fig.~\ref{fig:AlisLaw}.

Table~\ref{tab:halores} summarizes the results for the different
datasets and halo models. In particular, the Kepley et al. (2007)
dataset is consistent with a spherical dark matter halo. The Smith et
al. (2009a) dataset is consistent with oblate dark halo models with a
flattening in the potential of $g \approx 0.85$, corresponding to a
flattening in the dark matter density contours of $\approx 0.7$. Both
results assume that there is no contribution from the thin disc
potential and so this analysis provides lower limits to the axis ratio
$q$.  We shall revisit the effects of the thin disc in
Section~\ref{sect:StH}.

\begin{table}
\begin{center}
\renewcommand{\tabcolsep}{0.2cm}
\renewcommand{\arraystretch}{0.5}
\begin{tabular}{| l | c | c |}
    \hline 
  & Short Disc & Long Disc\\
    \hline
    \\
$b$ & [0.92,2.25] & [0.25,0.56] \\
$d$ & [0.49,0.81] & [0.78, 0.95] \\
$\mu$ & [0.3,0.7] & [0.7,1.4] \\
    \hline
  \end{tabular}
  \caption{Parameters for the the disc-halo decompositions adopted in
    Section~\ref{sec:TD} assuming that the dark matter potential has a
    logarithmic profile.}
\label{tab:parameters}
\end{center}
\end{table}

\subsection{The Dual Halo}

Even though the result given in words in eq~(\ref{eq:words}) is a
simple one, it can still be made to do some work. Here, we shall show
that it implies that the dual halo structure advocated by Carollo et
al. (2007, 2010) is unphysical.

Carollo et al. (2007) first argued for a two component halo on the
basis of different density profiles, metallicities and stellar orbits.
They claimed evidence for the existence of a metal-rich, inner halo
([Fe/H] $\sim$ -1.6) and a metal-poor outer halo ([Fe/H] $\sim$ -2.2)
The inner halo is supposed to have a modest net prograde rotation,
together with an oblate shape with an axis ratio of $q \sim 0.6$.  The
outer halo has net retrograde rotation with a much more spherical
distribution with an axis ratio of $q \sim 0.9$ to 1.0.

Carollo et al. (2010) supplied further structural and kinematical
analysis of the dual halo. The inner halo has a density profile that
falls off like $\rho \sim r^{-3.17 \pm 0.20}$.  It has velocity
dispersions $(\sigma_{RR}, \sigma_{\phi\phi}, \sigma_{zz}) = (150 \pm
2, 95 \pm 2, 85 \pm 2)$ kms$^{-1}$ around means of $(\bar v_R,
\bar v_\phi, \bar v_z ) = ( 3 \pm 2, 7 \pm 4, 3 \pm 1$)
kms$^{-1}$. This gives $K_{RR}/ K_{zz} = 4.40.$ for the inner halo.

The outer halo has a density profile that falls off like $\rho\sim
r^{-1.79 \pm 0.29}$. It has velocity dispersions $(\sigma_{RR},
\sigma_{\phi\phi}, \sigma_{zz}) = (159 \pm 4, 165 \pm 9, 116 \pm 3)$
kms$^{-1}$ around means of $(\bar v_R, \bar v_\phi, \bar v_z ) = ( -9
\pm 6, -80 \pm 13, 2 \pm 4$) kms$^{-1}$. This gives $K_{RR}/ K_{zz} =
4.40,$ for the outer halo, surprisingly close to the value obtained
for the inner halo.

Given that the fractional pressure support is the same for the inner
and outer halo, we see that the claimed flattenings and power-law
fall-offs are inconsistent with the virial theorem. If the density
profile of the inner halo declines more quickly than that of the outer
halo, then it cannot be flatter as well. Put simply, the claimed
shapes, profiles and kinematics of the inner and outer halo are
inconsistent with the structures moving in the same steady-state
potential.  If the outer halo is rounder than the inner halo, its
density profile must fall off more quickly than the inner one.

It is useful to summarise the assumptions in this conclusion.  The
result does assume that $q$, the flattening of the stellar halo, is
constant. There is good evidence for this as investigations of changes
of ellipticity with radius have typically found none (e.g., Sluis \&
Arnold 1998; Sesar et al. 2011; Deason et al. 2011b). The result also
assumes that $g$, the flattening of the dark matter equipotentials, is
constant. Whilst there is no direct evidence for this, it certainly
seems reasonable to investigate such simple models first of all.
Although we have used the asymptotic formula (\ref{eq:words}) in our
argument, the problem for Carollo et al.'s model becomes worse if the
full result is used. As can be seen from the curves in
Fig.~\ref{fig:power}, a greater pressure excess is needed at larger
flattenings than implied by a linear scaling with ellipticity.

\begin{table}
\begin{center}
\renewcommand{\tabcolsep}{0.2cm}
\renewcommand{\arraystretch}{0.5}
\begin{tabular}{| l l | l | l | l |}
  \hline 
  $\RDM$  & $c$ &  $\Rthin$ & $\mu$ & $\chi^{2}$ \\  
   (kpc) & \null & (kpc) & \null & \null  \\  
  \hline
  7 & 9.0 & 2.5 & 10.5 & 1.41\\
  7 & 7.3 & 2.6 & 8.0 & 1.41\\
  10 & 4.9 & 2.3 & 8.0 & 1.57\\
  10 & 4.7 & 2.5 & 8.0 & 1.37\\
  10 & 6.2 & 3.0 & 3.6 & 1.55\\
  12 & 3.7 & 2.5 & 8.1 & 1.37\\
  12 & 3.8 & 2.5 & 7.5 & 1.42\\
  12 & 4.7 & 3.0 & 2.8 & 1.32\\
  40 & 0.9 & 2.5 & 3.0 & 1.42\\
  40 & 1.5 & 3.0 & 1.1 & 1.39\\
  40 & 2.0 & 4.5 & 0.6 & 1.50\\
  200 & 0.1& 2.4 & 1.10 & 1.46\\
  200 & 0.1 & 2.8 & 0.28 & 1.29\\
  200 & 0.2& 3.1 & 0.20 & 1.45\\
  200 & 0.2 & 3.5 & 0.15 & 1.50\\
  \hline
  \end{tabular}
  \caption{Sample parameter values for the decompositions with an
    exponential disc with scalelength $\Rthin$ and NFW potential with
    scalelength $\RDM$ and concentration $c$ . The rotation curves are
    shown in Fig.~\ref{fig:rotcurves}. All the rotation curve fits
    take the surface density of the thin disc as 56 $M_\odot$
    pc$^{-2}$ at the solar radius of 8.0 kpc.}
\label{tab:nfw}
\end{center}
\end{table}
\begin{figure}
        \centering
 \includegraphics[width=0.45\textwidth]{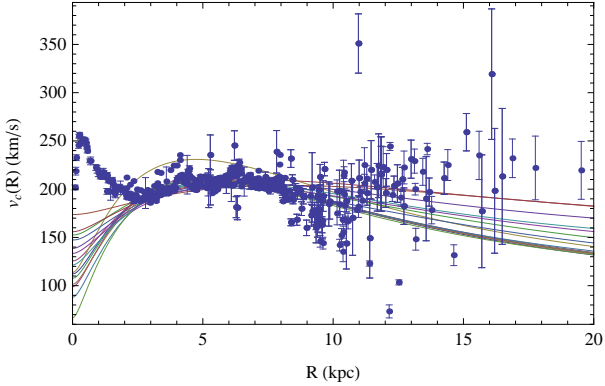}
 \caption{The rotation curves of model galaxies with NFW haloes and
   exponential thin discs with the parameters listed in
   Table~\ref{tab:nfw}.  The data are taken from Sofue et
   al. (2009). The fit is performed using the datapoints beyond 3 kpc,
   for which the contribution of non-circular motions caused by the
   bar is insignificant.}
\label{fig:rotcurves}
\end{figure}

\section{Thick Discs}

\label{sec:TD}

Our methods are also applicable to thick discs, and can give some
simple results. As thick discs are predominantly rotationally
supported, it makes sense to consider the total kinetic energy rather
than the pressure contribution alone.  Mass ordering arguments
suggest that we discard the contribution of the bulge in the virial
relations, even if it is important in shaping the innermost parts of
the rotation curve.

\begin{figure}
        \centering
 \includegraphics[width=0.45\textwidth]{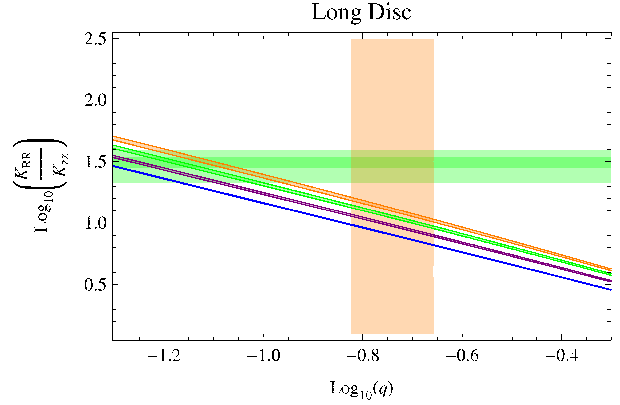}
 \caption{Virial sequences for the thick disc.  The equatorial to
   vertical kinetic energy ratio is shown as a function of the
   flattening $q$. The thick disc stars move in the potential of a
   long thin disc and logarithmic halo with relative mass ratio $\mu =
   0.7$. Each diagonal stripe corresponds to fixing $b = \Rthick/\RDM$
   and varying $d = \Rthin/\Rthick$ as in
   Table~\ref{tab:parameters}. The vertical and horizontal bands
   indicate the observationally preferred values of the flattening and
   the kinetic energy ratio.  Notice that all the models lie well
   below the observational window.  [Blue curves are for $b=0.2$,
     purple for $b = 0.25$, green for $b=0.3$ and orange for $b =
     0.35$. If $\mu$ is increased (decreased), the virial sequences
     move down (up) in the figure.]}
\label{fig:Long}
\end{figure}
\subsection{Formalism}

The density profiles of galactic discs tend to be exponential or
Gaussian -- more complicated than the homogeneous functions used to
represent stellar haloes.  We shall assume that the thick disc has a
double-exponential form:
\begin{equation}
\rho = \rho_0 \exp \left( - {R\over \Rthick } \right) \exp \left( -
  {|z| \over z_{\rm h}} \right)
\label{eq:thickdisc}
\end{equation}
and so the vertical scaleheight $z_{\rm h} = q \Rthick$, where $q$ is
as before the axis ratio of the density contours.

Thick discs move in the potential of both the dark halo and thin
disc. The latter can be represented by a
razor-thin\footnote{\change{We have calculated the correction to our
    virial terms by discarding the assumption of zero-thickness of the
    thin disc. We find it to be always $ < 7$ \% if the thin disk
    scaleheight is $<$ 350 pc for realistic values of the local
    surface density ($0.3 < \mu < 0.7$).}} exponential disc with
scale-length $\Rthin,$ and thus with potential:
\begin{equation}
  \Phi_{\rm disc}(R,z)=-2\pi G \Rthin^{2}\Sigma_{0}\int_{0}^{\infty}
  {\frac{J_{0}(kR)\mathrm{e}^{-k\left|
          z\right|}}{[1+\Rthin^2k^{2}]^{3/2}}}\mathrm{d}k.
\label{eq:thindisc}
\end{equation}
The ratio of the kinetic energy tensor components takes the form
\begin{equation} 
  {K_{RR} \over K_{zz}} = 
{\displaystyle \int{\rho
  R \partial_{R}(\Phi_{\rm disc} + \Phi_{\rm halo})\mathrm{d}^{3}x }
\over 
{\displaystyle \int{\rho z \partial_{z}
(\Phi_{\rm disc} + \Phi_{\rm halo})\mathrm{d}^{3}x }}},
\label{eq:defagain}
\end{equation}
and so depends on both the thin disc and halo potentials. By inserting
the thick disc density (\ref{eq:thickdisc}) and the thin disc
potential (\ref{eq:thindisc}) into (\ref{eq:defagain}), we can find
that the disc contributions are
\begin{eqnarray}
  D_{R}&\propto& 
{\displaystyle \int{\rho R \partial_{R}\Phi_{\rm disc} \mathrm{d}^{3}x }}\nonumber\\
 &=& \int_{0}^{\infty}{\frac{3k^{2}}{(1+qk)(1+k^{2})^{5/2}(1+k^{2}d^{2})^{3/2}}}\mathrm{d
  }k\ ,\\
  D_{z} &\propto&
{\displaystyle \int{\rho z \partial_{z}\Phi_{\rm disc} \mathrm{d}^{3}x
}}\nonumber \\
& =&\int_{0}^{\infty}{\frac{k}{(1+qk)^{2}(1+k^{2})^{3/2}(1+k^{2}d^{2})^{3/2}}}\mathrm{d
  }k\ ,
\end{eqnarray}
where $d = \Rthin/ \Rthick$. The choice of a double-exponential disc
enables the quadruple integrals to be reduced to a single one using
formulae [6.623] of Gradshteyn \& Ryzhik (1965). This simplification
is not possible for general thick-disc density profiles.

Now, let us suppose that the halo is the spherical limit of the
logarithmic model
\begin{equation}
\Phi_{\rm halo}=\Phi_{0}\log\left(r^{2} + \RDM^2\right).
\label{eq:sphbin}
\end{equation}
Then the halo contributions are
\begin{eqnarray}
H_{R} &=& 
b\int_0^\infty \mathrm{d} \chi \int_0^\infty \mathrm{d} \xi 
{\chi^3 \exp-(\chi+\xi)\over \chi^{2}+q^{2}\xi^{2}+b^{-2}},\\
H_{z} &=& 
bq \int_0^\infty \mathrm{d} \chi  \int_0^\infty \mathrm{d} \xi 
{\chi\xi^2 \exp-(\chi+\xi)\over \chi^{2}+q^{2}\xi^{2}+b^{-2}}
\end{eqnarray}
where $b = \Rthick/\RDM$. With some careful algebra, we can recast
eq.~(\ref{eq:defagain}) as
\begin{equation}
{K_{RR} \over K_{zz}} =   {1\over q} {2H_R + \mu D_R
    \over 2H_z + \mu D_z}\ ,
\end{equation}
where the constant $\mu$ is
\begin{equation}
\mu = {2\pi\Sigma_0 \Rthin^2 \over \Phi_0 \RDM}\ .
\end{equation}
In this case, the flattening is not fixed and there is no simple
dependence as was the case in the previous sections.  The flattening
in the total potential is controlled by the $\mu$ parameter, which
mediates the contributions of the razor-thin disc and the round dark
halo.

As an alternative model for the dark matter potential, we also examine
the spherical NFW profile, for which the gravitational force component
is
\begin{eqnarray}
\partial_{r}\Phi_{\rm halo} &=& \frac{4\pi
  G\rhoNFW \RDM^{3}}{r^{2}}
\left[\log\left(1\!+\!\frac{r}{\RDM}\right)-\frac{r}{r\!+\!\RDM}\right],
  \nonumber\\
  &=& \frac{4\pi G \rhoNFW\RDM^{3}}{r^{2}}\tilde{\Phi}(r/\RDM).
\end{eqnarray}
The disc contributions remain the same, but the halo ones become
\begin{eqnarray}
  H_{R} &=& 
  \int_0^\infty\!\mathrm{d}\chi\!\int_0^\infty \mathrm{d}\xi\! 
  {\chi^3 \exp\!-\!(\chi\!+\!\xi)\over (\chi^{2}+q^{2}\xi^{2})^{3/2}}
  \tilde\Phi(\argu)\  ,\\
  H_{z} &=& 
  q\!\int_0^\infty\!\mathrm{d}\chi\!\int_0^\infty\mathrm{d}\xi\!
  {\chi\xi^2\!\exp\!-\!(\chi\!+\!\xi)\over (\chi^{2}+q^{2}\xi^{2})^{3/2}}\tilde\Phi(\argu)\ ;
\end{eqnarray}
and the constant $\mu$ is now
\begin{equation}
\mu ={\Sigma_0 \Rthin^2 \over \rhoNFW \RDM^3}\ .
\end{equation}

In the limit of a massless thin disc $\mu \rightarrow 0$, the ratio
$K_{RR} / K_{zz}$ is roughly $\propto q^{-2}$ in the highly flattened
limit. This can be deduced from eq.~(\ref{eq:highlyflat}) using the
conjugateness theorem.
%
%
Letting the disc become massive, or equivalently introducing a
non-vanishing $\mu$, then the rise in pressure deficit for small $q$
is shallower, approximately $\propto q^{-1}$. This is understandable
because the presence of a thin disc flattens the total potential, and
thus a smaller pressure deficit is required to attain a given tracer
flattening when a disc is present.}

\begin{figure*}
        \centering
\includegraphics[width=0.45\textwidth]{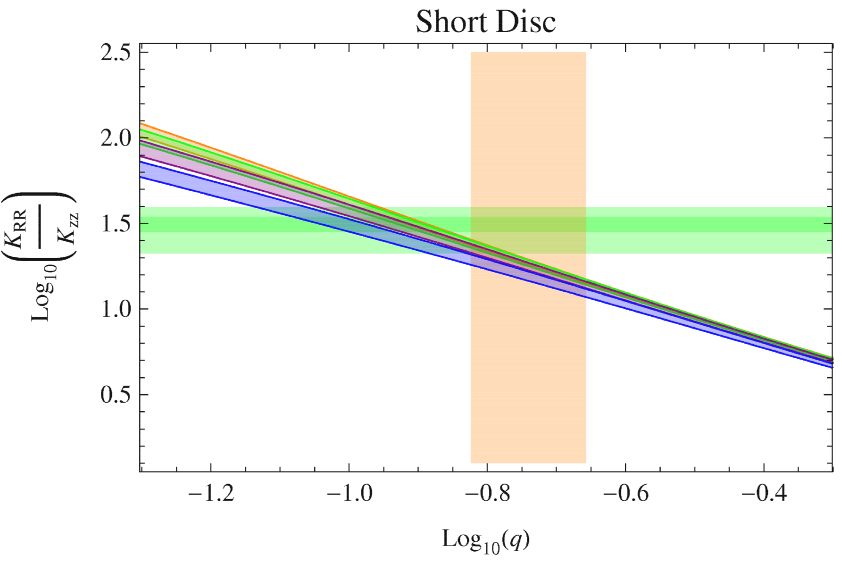}
\includegraphics[width=0.45\textwidth]{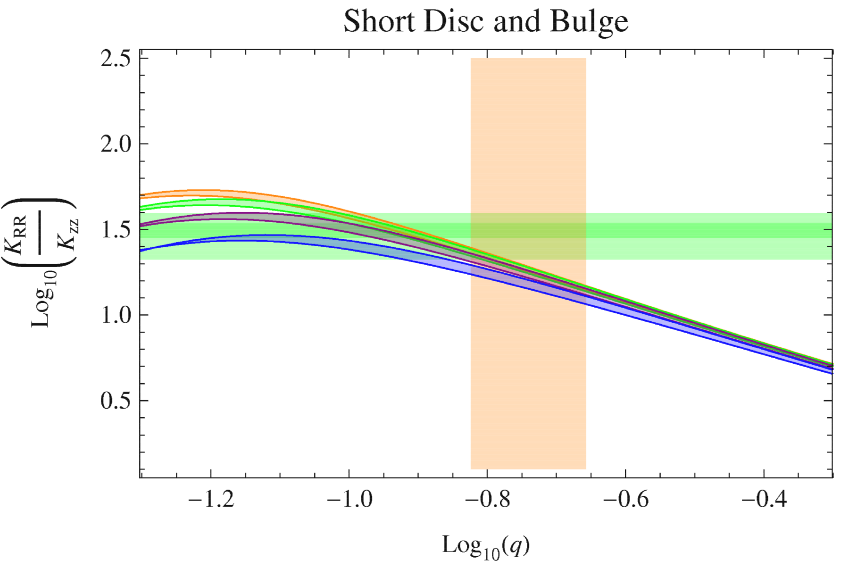}
\caption{Virial sequences for the thick disc.  Left: The thick disc
  resides in the potential of a short thin disc and logarithmic halo
  with relative mass ratio $\mu = 0.7$. Each diagonal stripe
  corresponds to fixing $b = \Rthick/\RDM$ and varying $d =
  \Rthin/\Rthick$ as in Table~\ref{tab:parameters}.  Now all the
  models cross the observational window, which is the intersection of
  the vertical orange and horizontal green stripes. Right: The thick
  disc stars now move in the potential of a short thin disc, a Sersic
  bulge and a logarithmic halo. Notice that the effects of including
  the bulge are small in the flattening regime indicated by the
  observationally preferred orange vertical band.  [Blue curves are
    for $b= 1.35$, purple for $b = 1.63$, green for $b=1.91$ and
    orange for $b = 2.25$. If $\mu$ is increased (decreased), the
    virial sequences move down (up) in the figure.]}
\label{fig:Shorter}
\end{figure*}

\subsection{Parameter Values}\label{ssect:parameters}

To make progress, we will need to choose astrophysically motivated
values for our thick disc models. Here, we are limited by the fact
that the thick disc of our Galaxy has not been fully mapped out.  We
shall see that with the existing data, a rather wide range of
parameters is plausible a priori.

De Jong et al. (2010) used SEGUE photometry to find the scalelength of
the thick disc $\Rthick = 4.1\pm0.4$ kpc and the scaleheight $z_{\rm
  h} = 0.75 \pm 0.07$ kpc. This gives the flattening or axis ratio as
\begin{equation}
  q =\frac{0.75\pm0.07}{4.1\pm0.4}\approx0.18\times(1\pm0.19)\ .
\end{equation}
This result is supported by the survey of 34 edge-on galaxies by
\citet{Yo06} who found that thick discs typically have $q = 0.18 \pm
0.04$.

The velocity dispersions in the thick disc vary with radius, perhaps
in a roughly exponential manner~\citep{Lew89}. The radial variation
has not been observationally determined in full, but this does not
harm us here. What is important in virial arguments is the ratio of
the global averages of horizontal to vertical kinetic energy. This
ratio \change{can be} well constrained even though the variation of
velocity dispersion in the thick disc is not yet fully mapped out. In
particular, if the velocity dispersions are proportional to
$\sqrt{\rho}$ and therefore vary exponentially, then the ratio of the
local averages of horizontal to vertical kinetic energy is equal to
the global averages.~\footnote{There is some tentative evidence in
Figure 12 of \citet{Ca11} that the vertical velocity dispersion of the
thick disc does follow a different law as compared to the horizontal
velocity dispersions. \change{Nonetheless, the error bars on these
measurements remain quite large (see e.g. Table 3 of their paper).}}
This provides some security that using comparatively local
measurements of the thick disc velocity dispersions leads to
meaningful kinetic energy ratios.

Carollo et al. (2010) explored a local volume within 4 kpc of the Sun
and reported that the velocity ellipsoid of the thick disc is
$(\sigma_{RR}, \sigma_{\phi\phi}, \sigma_{zz}) = (53 \pm 2, 51 \pm 2,
35 \pm 1)$ kms$^{-1}$ around means of $(\bar v_R, \bar v_\phi, \bar
v_z ) = ( 3 \pm 2, 182 \pm 2, 0 \pm 1$) kms$^{-1}$.  This gives
$K_{RR}/ K_{zz} = 31.5\times(1\pm0.06)$.  There are also values
available from the work of \citet{Ca11}, who used red clump stars
between 5 and 10 kpc of the Galactic Center and between 1 and 3 kpc
from the Galactic plane.  Stars closer to the Galactic plane are not
used to minimize contamination from the thin disc.  For the thick
disc, they found $(\sigma_{RR}, \sigma_{\phi\phi}, \sigma_{zz}) = (70
\pm 4, 48 \pm 8, 36 \pm 4)$ kms$^{-1}$ around a rotation velocity of
$\bar v_\phi = 175 \pm 2$ kms$^{-1}$.  This yields $K_{RR}/ K_{zz}
=29.2\times(1\pm0.23)$. Although the values of the dispersions found
in the two investigations are not the same, the pressure ratio is much
more robust, which adds weight to our argument that the global ratio
is well constrained.

The halo and thin disc parameters are of course strongly correlated,
as the combined force field must generate the Galactic rotation
curve. Sofue et al. (2009, 2010) show that a decomposition into a de
Vaucouleurs bulge, exponential disc with spiral arms, and isothermal
dark halo gives values of the halo core radius $\RDM \approx 8 - 15$
kpc.  Sofue et al.'s thin disc has a local surface density of $87
M_\odot {\rm pc}^{-2}$ and a scalelength $\Rthin = 3.5$ kpc --
somewhat larger than is found by most other
investigators~\citep[e.g.,][]{Fu94, Oh01, Ca11} -- and accordingly we
refer to it as the {\it long disc} model. 

On the other hand, \citet{Ho04} used Hipparcos data to measure the
local disc surface density as $56 \pm 6 M_\odot {\rm pc}^{-2}$, lower
than that estimated by \citet{So10}.  \citet{Ca11} used kinematical
and photometric data to infer the thin disc scale length as $\Rthin =
2.6 \pm 0.4$ kpc. These observations suggest that the thin disc may be
less massive and smaller than advocated by \citet{So09}, and
accordingly we refer to this as the \textit{short disc} model. To fit
the rotation curve, this latter model requires lower values of $\RDM$,
accordingly in the range $\approx 1.5 - 3$ kpc. Of course, this is an
underestimate of the true scalelength of the dark matter, as we have
not explicitly included a bulge. With these two choices (short versus
long disc), the ratios $b$, $d$ and $\mu$ are allowed to vary in the
ranges summarized in Table~\ref{tab:parameters}.  Under the assumption
of a logarithmic halo, the mass ratio $\mu$ in the range $\approx 0.7
- 1.4$ (long disc) and $\approx 0.3 - 0.7$ (short disc) to provide a
reasonable match to the rotation curve. Similarly, the ranges of $b =
\Rthick/\RDM$ are constrained to lie in the range $0.25 \lesssim b
\lesssim 0.56$ (long) and $0.92 \lesssim b \lesssim 2.25$
(short). From the scalelengths of the thin and thick discs, we can
deduce the range of $d = \Rthin/ \Rthick$ as $0.78 \lesssim d \lesssim
0.95$ (long) and $0.49 \lesssim d \lesssim 0.81$ (short).

Matters are more complicated if a NFW profile is chosen for the dark
matter potential. Here, we fit the rotation curve data of \citet{So09}
with an exponential disc and NFW halo, as shown in
Fig.~\ref{fig:rotcurves}.  Even keeping the local disc density as $56
M_{\odot} {\rm pc}^{-2}$ and $\Rthin\gtrsim 2.4$ \rm kpc, there is
quite a large range in which the other parameters may vary. We list
the parameters of the models in Table~\ref{tab:nfw}, although we shall
show that their virial sequences are not too different.

\begin{figure}
        \centering
\includegraphics[width=0.45\textwidth]{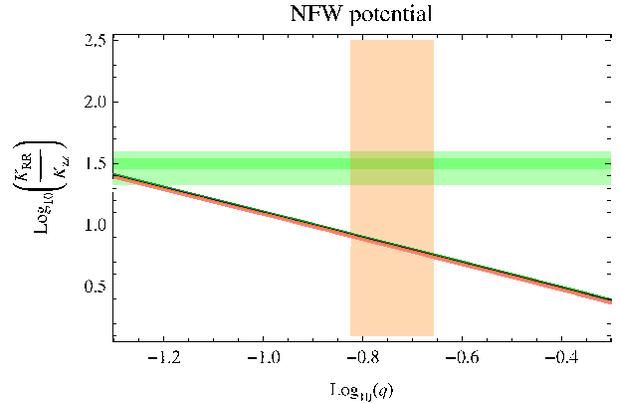}
\caption{As Fig.~\ref{fig:Shorter}, but for a NFW rather than the
  logarithmic halo. Plotted are all the virial sequences for the
  models with rotation curves displayed in
  Fig.~\ref{fig:rotcurves}. Different colours stand for different
  values of $b = \Rthick/ \Rthin$, but the sequences are so near to
  one another that listing the colour correspondence is superfluous.}
\label{fig:NFW}
\end{figure}

\subsection{Short versus Long Thin Discs}

Our method is to take $\Rthick = 4.1 \pm 0.4$ kpc from de Jong et
al. (2010) and $\Rthin$ either short at $2.6 \pm 0.4$ kpc
(Casetti-Dinescu et al (2011) or long at 3.5 kpc from Sofue et al
(2009). Then a fit to Sofue's data -- as judged by the likelihood
falling to half its maximum value -- is used to find plausible ranges
of $d$ and $\mu$ . Virial sequences are found by fixing $b$, and $\mu
= 0.7$, then varying $q$ to get a single line across the plane, and
finally $d$ (within the range found from the rotation curve) to get a
band of finite width rather than a single line.

The panels of Figs~\ref{fig:Long} and \ref{fig:Shorter} show such
virial sequences. They give the thick disc kinetic energy ratio as a
function of flattening $q$. The allowed observational range is shown
by vertical and horizontal stripes, so that the intersection gives the
region compatible with the data. Specifically, the vertical orange
stripe corresponds to the flattening of the thick disc inferred by
\citet{De10}, whilst the horizontal green stripes correspond to the
kinetic energy ratios from \citet{Ca10} and \citet{Ca11}.

Plotted on Figs~\ref{fig:Long} and ~\ref{fig:Shorter} are the virial
sequences of the long and short discs respectively.  We see that the
short disc does indeed give a better match to the observations than
the long disc, even when the $\mu$ parameter is taken towards the high
end of its possible range.  By contrast, the virial sequences for the
long disc yield pressure ratios that lie appreciably below the
estimated ones, for all the relevant values of $\mu$ parameter.

The effects of incorporating a bulge are shown in the right panel of
Fig~\ref{fig:Shorter}. The fit to the rotation curve is performed
again, using a short thin disc, an isothermal dark halo and a
spherical Sersic $m=3$ bulge, given by the deprojection of
\begin{equation}
\Sigma(R) = \Sigma_0 \exp \left( - \kappa (R/R_{\rm e})^{1\over3} \right)
\end{equation}
Here, $\Sigma$ is the projected bulge density on the sky, $R$ is
projected radius and $R_{\rm e}$ is the effective radius and $\kappa$
is a constant. The values of the parameters are chosen to give a total
bulge mass of $\sim 10^{10} M_\odot$ and a maximum contribution to the
rotation curve of $\sim 100$ kms$^{-1}$. The parameter values are
suggestive rather than definitive, as we merely wish to ascertain the
importance of the effect.  In fact, we notice that the virial
sequences do not change much in the observational regime -- the left
and right panels of Fig.~\ref{fig:Shorter} are very similar for $q
\gtrsim 0.1$. This is understandable, as the models with or without a
bulge reproduce the same overall rotation curve, albeit using
different components. The virial quantities do not change much, as
they depend on the total potential.

We conclude that the flattening and kinematics of the thick disc are
consistent with stellar motion in a potential generated by a thin disc
with a short scalelength $\sim 2.6$ kpc, together with the
normalisation suggested by \citet{Ho04}. In particular, thin discs
with longer scalelengths $\sim 3.5$ kpc are necessarily more massive
if they are to reproduce the local surface density. The stronger
gravity field towards the equatorial plane then causes the thick disc
to be flatter than is measured, given the observed kinematics.

We briefly show the differences made by using a NFW halo rather than a
logarithmic one in Fig.~\ref{fig:NFW}. Here, all the virial sequences
corresponding to the parameters in Table~\ref{tab:nfw} lie below the
observational estimates for flattening versus pressure excess. The
reason for this is as follows. For cusped potentials, the virial terms
are dominated by contributions from near the central cusp. Near the
very centre, the combined equipotentials due to the thin disc and
spherical halo are flatter than the tracer density. If this region is
weighted more in the virial integrals, then the average $K_{zz}$
required to produce the flattening is greater and so the virial
sequences of an NFW halo fall below those of a logarithmic halo and
also the observational data (cf.  Figs.~\ref{fig:Shorter} and
\ref{fig:NFW}). Thus, a logarithmic potential with a finite core seems
to be favoured over an NFW profile, at least as judged from the
present constraints on the thick disc kinematics.

\begin{figure}
        \centering
\includegraphics[width=0.45\textwidth]{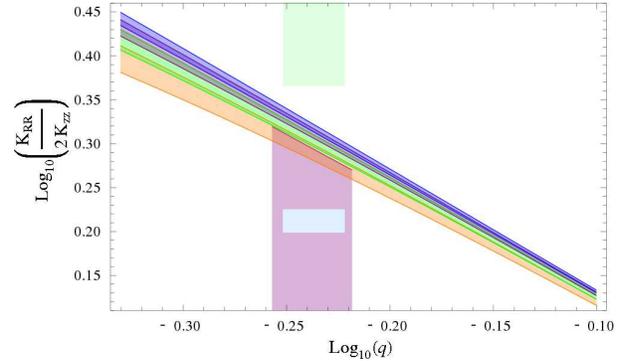}
\caption{Virial sequences for the stellar halo in a spherical
  logarithmic dark halo plus a short thin disc. The relative weighting
  of the disc and halo is controlled by $\mu = 0.7$. Each virial
  sequence corresponds to fixing $\tb = \RB/ \RDM$ and varying $\td
  =\Rthin/ \RB$ from 0.09 to 0.12 (see text).  The models all lie
  above the pale blue \citet{Sm09} databox.  Lower values of $\mu$
  yield somewhat higher pressure ratios, but still below the pale
  green \citet{Ke07} databox. The purple shaded zone shows models with
  constant anisotropy and spherical alignment, with the extreme radial
  orbit model providing the diagonal edge. [Blue curves are for $b=7$,
  purple for $b = 8.6$, green for $b=10.2$ and orange for $b = 12$.]}
\label{fig:StHdDM}
\end{figure}

%

\section{The Stellar Halo Revisited}
\label{sect:StH}

In Section~\ref{sec:apps}, we presented a simple model of the stellar
halo, and showed that if the local measurements of the velocity
dispersion using the SDSS subdwarfs (Smith et al. 2009a) are a good
guide to the global kinematics, then this implies a moderately
flattened potential with an axis ratio of $g \approx 0.85$. If this is
ascribed solely to the dark matter halo, then it corresponds to a
flattening in the dark matter density contours of $\approx 0.7$.

In Section~\ref{sec:TD}, we showed that the global kinematics of the
thick disc are consistent with motion in the combined potential of a
spherical dark halo together with a thin disc.  Here, we will ask: is
it possible that the dark halo is spherical, and that the flattening
in the potential measured from the kinematics of halo stars is due
entirely to the presence of the thin disc?  If the dark halo is round,
then the flattening of the combined disc and halo equipotentials
changes from $\approx 0.5$ close to the Galactic centre to $\approx 1$
at Galactocentric distances beyond 10 kpc.

The gravity field is now provided by the thin disc potential with
short scalelength $\sim 2.6$ kpc, together with a completely spherical
logarithmic dark matter potential as in eq~(\ref{eq:sphbin}).  The
stellar halo density is given by the broken power-law of
eq~(\ref{eq:densAlis}) with break radius $\RB$. The two lengthscales
involved are
\begin{equation}
\tb=\frac{\RB}{\RDM}\qquad\qquad \td=\frac{\Rthin}{\RB},
\label{eq:newlengthscales}
\end{equation}
We compute virial sequences for $\tb$ in the range 7 to 14 and $\td$
in the range 0.09 to 0.12, which are suggested by the roughly known
values of the scalelengths (that is, $\Rthin = 2.6 \pm 0.4$ kpc, $\RB
\approx 2$ - $4$ kpc and $\RB \approx 27$ kpc)

Fig.~\ref{fig:StHdDM} shows the pressure ratio $\Pi_{RR}/\Pi_{zz}$ as
a function of $q$, the flattening of the stellar halo.  The light blue
and green boxes indicate the observational regimes, depending on
whether the pressure ratio is estimated from the data of \citet{Sm09}
or of \citet{Ke07} assuming cylindrical alignment.  In each case, the
constraints on flattening come from the work of \citet{De11b}.  The
virial sequences lie below the \citet{Ke07} databox and above the
\citet{Sm09} databox.  Accordingly, it seems reasonable to argue that
the underlying assumption of a spherical dark halo may hold good, as
the models are not too far off the uncertain data. The purple zone
covers the parts of the ($\Pi_{RR}/\Pi_{zz}, q$) plane for which the
velocity ellipsoid is spherically aligned and the anisotropy is
constant using eq~(\ref{eq:anisotropy}). The upper diagonal line
corresponds to the purely radial orbit model. Hence, the models that
do reproduce the observations are radially biased with an anisotropy
parameter $\beta \approx 0.9$. This makes good sense, as the
underlying dark halo is nearly spherical at large radii where the disc
becomes unimportant, in which case the flattening can only be produced
by substantial radial anisotropy. However, such radial anisotropy is
much more extreme than implied by Kepley et al.'s (2007) or Smith et
al's (2009a) data, which have $\beta \approx 0.6$.


\section{Conclusions}

The tensor virial theorem can provide a powerful technique for
extracting gross structural properties, such as flattenings, from
datasets. It has been widely used in galactic astronomy, particularly
for self-gravitating systems -- such as stars in elliptical galaxies
--- moving in the potential generated by their density field. The
literature on applications to tracer populations, in which the stars
are primarily moving in the gravity field provided by dark matter or
other stellar components, is quite sparse~\citep{Wh89,So89,He08}.

We have provided the first systematic exploitation of the tensor
virial theorem for tracer stellar populations, and shown how it can be
used to yield constraints on global kinematics and flattening. This
technique offers considerable advantages over, for example, the Jeans
equations. The latter depend on the derivatives of densities and
velocity dispersions, and are necessarily subject to uncertainties
with noisy data. The tensor virial theorem deals with integrated bulk
quantities, which are less susceptible to observational errors or
systematic problems such as substructure in the data.

An obvious application is to Population II stars moving in the dark
halo of our Galaxy.  For spheroidal tracer populations with flattening
$q$ moving in spherical power-law potentials, there are analytic
results that prescribe the ratio of globally averaged equatorial to
vertical velocity dispersion. Thanks to the Flattening Theorem, the
results can be straightforwardly adapted to spheroidal dark halos with
axis ratio $g$ in the equipotentials. The virial quantities depend
only on the ratio $q/g$ in this instance. 


The main astrophysical results of the paper are threefold. First, we
have shown how to interpret the stellar kinematical measurements of
halo stars in terms of the global flattening of the dark halo. In
particular, if the Smith et al. (2009a) values for the velocity
dispersion of the halo subdwarfs are accepted, then the dataset is
consistent with oblate dark halo models with a flattening in the
potential of $g \approx 0.85$, corresponding to a flattening in the
dark matter density contours of $\approx 0.7$; this is driven by the
radial anisotropy that Smith et al. found. If however, the Kepley et
al. (2007) values are preferred, then the dark halo may well be close
to spherical. We have to make the assumption that the local kinematics
are proxies for the global kinematics, which is assuredly a weakness
in such analyses at present. However, the GAIA satellite, which is
scheduled to fly in 2013, will for the first time provide us with
detailed information on the three-dimensional kinematics of halo
stars, and so there is hope for real advances using this technique in
the near future.

Secondly, we have demonstrated that the dual halo structure as
proposed by Carollo et al. (2010) appears to be incorrect.
It is not possible to support an inner halo and an outer halo with the
claimed velocity dispersions in the same gravitational potential: the
model is inconsistent with the requirements of the virial theorem.  A
powerful aspect of virial methods is the fact that multiple
populations residing in the same potential can be used to
cross-constrain the dark hark halo mass and shape, without the need of
building detailed models to reproduce the photometry and kinematics.
There is considerable scope for extending our virial techniques to
study multiple populations moving in the same gravity field, a topic
that is already important in studies of the dwarf
spheroidals~\citep{Am11, Wa11}.

Thirdly, we have carried out an investigation into the kinematics of
the thick disc.  We find that the reported thick disc kinematics are
consistent with a thin disc with scalelength and an isothermal dark
halo. Specifically, we find that the thick disc kinematics and shape
is consistent with a potential generated by a thin disc with
scalelength $\approx 2.6$ kpc, and surface-density normalisation of
$56 \pm 6 M_\odot {\rm pc}^{-2}$ \citep{Ho04}. Although NFW potentials
can be used to fit the rotation curve, they seemingly fail to
reproduce the virial ratios expected for the thick disc, as judged
from the observational data.

The near future sees a number of ambitious radial velocity surveys,
such as ESO-GAIA and Hermes. These will complement the abundant
positional and proper motion data on stars in the Galaxy that will be
delivered by the GAIA satellite.  Given the quality and quantity of
the datasets, it is an attractive proposition to look at virial
quantities to gain insight into the Galaxy's potential, perhaps as a
prelude to detailed phase space modelling.  We anticipate that the
tensor virial theorem will become an important tool to make sense of
the very large datasets on stellar positions and velocities that will
become available in the future.

\section*{Acknowledgments}
AA thanks the Science and Technology Facilities Council and the Isaac
Newton Trust for the award of a studentship. We acknowledge useful
discussions with Jin An, Nicola Amorisco, Vasily Belokurov, Alis
Deason, Donald Lynden-Bell and Martin Smith. We wish to thank an
anonymous referee who considerably improved the paper with comments
and suggestions.

\label{lastpage}

\begin{thebibliography}{99}

\bibitem[Amorisco \& Evans (2011)]{Am11} Amorisco N., Evans N.W.,
  2011, MNRAS, in press.

\bibitem[An et al.(2004)]{2004MNRAS.351.1071A} An, J.~H., et al.\ 2004, 
\mnras, 351, 1071 

\bibitem[Battaglia et al.(2005)]{Ba05} Battaglia, G., Helmi, 
A., Morrison, H., et al.\ 2005, \mnras, 364, 433 

\bibitem[Bell et al.(2008)]{2008ApJ...680..295B} Bell, E.~F., et al.\ 2008, 
\apj, 680, 295 

\bibitem[Binney(1978)]{Bi78} Binney, J.\ 1978, \mnras, 183, 
501 

\bibitem[Binney(1981)]{1981MNRAS.196..455B} Binney, J.\ 1981, \mnras, 196, 
455 

\bibitem[Binney(2005)]{Bi05} Binney, J.\ 2005, \mnras, 363, 
937 

\bibitem[Bramich et al.(2008)]{2008MNRAS.386..887B} Bramich, D.~M., et al.\ 
2008, \mnras, 386, 887 

\bibitem[Carollo et al.(2007)]{Ca07} Carollo, D., et al.\ 
2007, \nat, 450, 1020 

\bibitem[Carollo et al.(2010)]{Ca10} Carollo, D., et al.\ 
2010, \apj, 712, 692 

\bibitem[Casetti-Dinescu et al.(2011)]{Ca11} Casetti-Dinescu, D.~I.,
  Girard, T.~M., Korchagin, V.~I., \& van Altena, W.~F.\ 2011, \apj,
  728, 7

\bibitem[Chandrasekhar(1987)]{Ch87} Chandrasekhar S. \ 1987,
  Ellipsoidal Figures of Equilibrium, Dover

\bibitem[Ciotti \& Morganti(2010)]{2010MNRAS.408.1070C} Ciotti, L., \& Morganti, L.\ 2010, \mnras, 408, 1070;

\bibitem[de Jong et al.(2010)]{De10} de Jong, J.~T.~A., 
Yanny, B., Rix, H.-W., et al.\ 2010, \apj, 714, 663 

\bibitem[Deason et al.(2011a)]{De11a} Deason, A.~J., 
Belokurov, V., \& Evans, N.~W.\ 2011a, \mnras, 411, 1480 

\bibitem[Deason et al.(2011b)]{De11b} Deason, A.~J., 
Belokurov, V., \& Evans, N.~W.\ 2011b, \mnras, 416, 2903

\bibitem[Evans(1993)]{Ev93} Evans, N.~W.\ 1993, \mnras, 260, 
191 

\bibitem[Evans(1994)]{Ev94} Evans, N.~W.\ 1994, \mnras,
  267, 333

\bibitem[Evans et al.(1997)]{Ev97} Evans, N.~W., Hafner, 
R.~M., \& de Zeeuw, P.~T.\ 1997, \mnras, 286, 315 

\bibitem[Fux \& Martinet(1994)]{Fu94} Fux, R., \& Martinet, L.\ 1994,
  \aap, 287, L21

\bibitem[Gradshteyn \& Ryzhik (1965)]{Gr65} Gradshteyn I.S., Ryzhik
  I.M.  1965, Tables of Integrals, Series and Products, Academic
  Press, New York.

\bibitem[Helmi(2008)]{He08} Helmi, A.\ 2008, \aapr, 15, 145 

\bibitem[Holmberg \& Flynn(2004)]{Ho04} Holmberg, J., \& Flynn, C.\
  2004, \mnras, 352, 440

\bibitem[Kepley et al.(2007)]{Ke07} Kepley, A.~A., et al.\ 2007, AJ,
  134, 1579

\bibitem[Lewis \& Freeman(1989)]{Lew89} Lewis, J.~R., \& Freeman,
  K.~C.\ 1989, \aj, 97, 139

\bibitem[Navarro et al.(1996)]{1996ApJ...462..563N} Navarro, J.~F., Frenk, 
C.~S., \& White, S.~D.~M.\ 1996, ApJ, 462, 563 

\bibitem[Ojha(2001)]{Oh01} Ojha, D.~K.\ 2001, \mnras, 322, 426

\bibitem[Petrou(1981)]{Pe81} Petrou, M., 1981, PhD Thesis, University
  of Cambridge

\bibitem[Roberts(1962)]{Roberts62} Roberts, P.H.\ 1962, \apj, 136,
  1108

\bibitem[S{\'e}rsic(1963)]{1963BAAA....6...41S} S{\'e}rsic, J.~L.\ 1963, 
Boletin de la Asociacion Argentina de Astronomia La Plata Argentina, 6, 41 

\bibitem[Sesar et al.(2011)]{2011ApJ...731....4S} Sesar, B., Juri{\'c}, M., 
\& Ivezi{\'c}, {\v Z}.\ 2011, \apj, 731, 4 

\bibitem[Smith et al.(2009a)]{Sm09} Smith, M.~C., et al.\ 2009a, \mnras,
  399, 1223

\bibitem[Smith et al.(2009b)]{Sm09b} Smith, M.~C., Evans, N.W., An,
  J.~H.\ 2009b, \apj, 698, 1110

\bibitem[Sluis \& Arnold(1998)]{1998MNRAS.297..732S} Sluis, A.~P.~N.,
  \& Arnold, R.~A.\ 1998, \mnras, 297, 732

\bibitem[Sofue et al.(2009)]{So09} Sofue, Y., Honma, M., \& Omodaka,
  T.\ 2009, \pasj, 61, 227

\bibitem[Sofue et al.(2010)]{So10} Sofue, Y., Honma, M., \& Omodaka,
  T.\ 2010, \pasj, 62, 1367

\bibitem[Sommer-Larsen \& Christensen(1989)]{So89} Sommer-Larsen, J.,
  \& Christensen, P.~R.\ 1989, \mnras, 239, 441

\bibitem[van der Marel(1991)]{Ma91} van der Marel, R.~P.\ 1991, MNRAS,
  253, 710

\bibitem[Walker \& Penarrubia (2011)]{Wa11} Walker M. G., Penarrubia J.,
2011, ApJ, in press

\bibitem[Watkins et al.(2009)]{Wa09} Watkins, L.~L., et al.\ 
2009, \mnras, 398, 1757 

\bibitem[Watkins et al.(2010)]{2010MNRAS.406..264W} Watkins, L.~L., Evans, 
N.~W., \& An, J.~H.\ 2010, \mnras, 406, 264 

\bibitem[White(1989)]{Wh89} White, S.~D.~M.\ 1989, MNRAS, 237, 51P

\bibitem[Yanny et al.(2009)]{Ya09} Yanny, B. et al.\ 2009, AJ, 137,
  4377

\bibitem[Yoachim \& Dalcanton(2006)]{Yo06} Yoachim, P., \& Dalcanton,
  J.~J.\ 2006, \aj, 131, 226

\end{thebibliography}
\end{document}